\font\twelverm = cmr10 scaled\magstep1 \font\tenrm = cmr10
       \font\sevenrm = cmr7
\font\twelvei = cmmi10 scaled\magstep1
       \font\teni = cmmi10 \font\seveni = cmmi7
\font\twelveit = cmti10 scaled\magstep1 
       
\font\twelvesy = cmsy10 scaled\magstep1
       \font\tensy = cmsy10 \font\sevensy = cmsy7
\font\twelvebf = cmbx10 scaled\magstep1 \font\tenbf = cmbx10
       \font\sevenbf = cmbx7
\font\twelvesl = cmsl10 scaled\magstep1
\font\twelvett = cmtt10 scaled\magstep1
\font\mbf = cmmib10 scaled\magstep1
       \font\mbfs = cmmib10 \font\mbfss = cmmib10 scaled 833
\font\msybf = cmbsy10 scaled\magstep1
       \font\msybfs = cmbsy10 \font\msybfss = cmbsy10 scaled 833

%
\textfont0 = \twelverm
       \scriptfont0 = \tenrm    \scriptscriptfont0 = \sevenrm
       \def\rm{\fam0 \twelverm}
\textfont1 = \twelvei
       \scriptfont1 = \teni    \scriptscriptfont1 = \seveni
       
\textfont2 = \twelvesy
       \scriptfont2 = \tensy    \scriptscriptfont2 = \sevensy
       
\newfam\itfam \def\it{\fam\itfam \twelveit} \textfont\itfam=\twelveit
\newfam\slfam  \textfont\slfam=\twelvesl
\newfam\bffam \def\bf{\fam\bffam \twelvebf} \textfont\bffam=\twelvebf
       \scriptfont\bffam = \tenbf    \scriptscriptfont\bffam = \sevenbf
\newfam\ttfam  \textfont\ttfam=\twelvett
\textfont9 = \mbf
       \scriptfont9 = \mbfs \scriptscriptfont9 = \mbfss
       \def\bmit{\fam9 }
\textfont10 = \msybf
       \scriptfont10 = \msybfs \scriptscriptfont10 = \msybfss
       \def\bmsy{\fam10 }
%
%
\rm
\hsize=6.5in
\hoffset=.0in
\vsize=9in
\voffset=0.7in
\baselineskip=24pt
%
%
\raggedright  \pretolerance = 800  \tolerance = 1100
\raggedbottom
%
%
\dimen1=\baselineskip \multiply\dimen1 by 3 \divide\dimen1 by 4
\dimen2=\dimen1 \divide\dimen2 by 2
%
%
\def\apjsingle{\baselineskip = 14pt
               \parskip=14pt plus 1pt
               \dimen1=\baselineskip \multiply\dimen1 by 3 \divide\dimen1 by 4
               \dimen2=\dimen1 \divide\dimen2 by 2
               \scriptfont0 = \tenrm  \scriptscriptfont0 = \sevenrm
               \scriptfont1 = \teni  \scriptscriptfont1 = \seveni
               \scriptfont2 = \tensy \scriptscriptfont2 = \sevensy
               \scriptfont\bffam = \tenbf \scriptscriptfont\bffam = \sevenbf
               \rightskip=0pt  \spaceskip=0pt  \xspaceskip=0pt
               \pretolerance=200  \tolerance=400
              }
%
%
\nopagenumbers
\headline={\ifnum\pageno=1 \hss\thinspace\hss
     \else\hss\folio\hss \fi}
%
%
\def\heading#1{\vskip \parskip  \vskip \dimen1
     \centerline{#1}
     \vskip \dimen2}
%
%
\count10 = 0
\def\section#1{\global\advance\count10 by 1
    \vskip \parskip  \vskip \dimen1
    \centerline{\number\count10.\ {#1}}
    \global\count11=0
    \vskip \dimen2}
%
%
\def\subsection#1{\global\advance\count11 by 1
    \vskip \parskip  \vskip \dimen1
    \centerline{\number\count10.\number\count11.\ {\it #1}}
    \global\count12=0
    \vskip \dimen2}
%
%
\def\subsubsection#1{\global\advance\count12 by 1
    \vskip \parskip  \vskip \dimen1
    \centerline{\number\count10.\number\count11.\number\count12.\ {\it #1}}
    \vskip \dimen2}
%
%
\def\refindent{\advance\leftskip by 24pt \parindent=-24pt \parskip=2pt}
%
%
\def\journal#1#2#3#4#5{{\refindent
                      {#1}        
                      {#2},       
                      {#3},       
                      {#4},       
                      {#5}.       
                      \par}}
%
%
\def\infuture#1#2#3#4{{\refindent
                  {#1}         
                  {#2},        
                  {#3},        
                  {#4}.        
                  \par }}
%
%

%
%

%
%
\def\book#1#2#3#4#5{{\refindent
                   {#1}         
                   {#2},        
                   {#3}         
                  ({#4}:        
                   {#5}).       
                   \par }}
%
%
\def\privcom#1#2#3{{\refindent
                  {#1}        
                  {#2},       
                  {#3}.       
                  \par }}
%
%

%
%

%
%
\def\figcap#1#2{{\refindent
                   Fig. {#1}.---   
                        {#2}       
                        \par}}
%
%
\def\monthchar{\ifcase\month
           \or January   \or February \or March    \or April
           \or May       \or June     \or July     \or August
           \or September \or October  \or November \or December\fi}
\def\today{\ifcase\month
           \or January   \or February \or March    \or April
           \or May       \or June     \or July     \or August
           \or September \or October  \or November \or December\fi
           \space\number\day, \number\year}

\def\etal{{\it et al.}}
\def\eg{{\it e.g.}}
\def\ie{{\it i.e.}}
\def\lsim{\raise0.3ex\hbox{$<$}\kern-0.75em{\lower0.65ex\hbox{$\sim$}}}
\def\gsim{\raise0.3ex\hbox{$>$}\kern-0.75em{\lower0.65ex\hbox{$\sim$}}}
\apjsingle
\centerline{~~~}
\vskip 2cm
\centerline{\bf Numerical Magnetohydrodynamics in Astrophysics:}
\centerline{\bf Algorithm and Tests for One-Dimensional Flow}
\vskip 3cm
\centerline{Dongsu Ryu}
\vskip 0.3cm
\centerline{Princeton University Observatory, Peyton Hall, Princeton,
NJ 08544;}
\centerline{Department of Astronomy and Space Science, Chungnam National
University,}
\centerline{Daejeon 305-764, (South) Korea}
\vskip 0.3cm
\centerline{and}
\vskip 0.3cm
\centerline{T.~W.~Jones}
\vskip 0.3cm
\centerline{School of Physics and Astronomy, University of Minnesota,
Minneapolis, MN 55455}
\vskip 4cm
\centerline{\today}
\vskip 3cm
\centerline{Submitted to the Astrophysical Journal}
\vfill\eject
\heading{ABSTRACT}
We describe a numerical code to solve the equations for
ideal magnetohydrodynamics (MHD).
It is based on an explicit finite difference scheme on an
Eulerian grid, called the Total Variation Diminishing (TVD)
scheme, which is a second-order-accurate extension
of the Roe-type upwind scheme.
We also describe a nonlinear Riemann solver for ideal MHD,
which includes rarefactions as well as shocks and produces
exact solutions for two-dimensional magnetic field structures as
well as for the three-dimensional ones.
The numerical code and the Riemann solver have been used to
test each other.

Extensive tests encompassing all the possible ideal MHD
structures with planar symmetries (\ie ~one-dimensional flows) are presented.
These include those for which the field structure is two-dimensional
({\it i.e.}, those flows often called ``$1 + 1/2$
dimensional'') as well as those for which the magnetic field plane
rotates ({\it i.e.,}, those flows often called ``$1 + 1/2 + 1/2$
dimensional'').
Results indicate that the code can resolve strong fast,
slow, and magnetosonic shocks within 2-4 cells while more cells
are required if shocks become weak.
With proper stiffening, rotational discontinuities are resolved
within 3-5 cells.
Contact discontinuities are also resolved within 3-5 cells with
stiffening and 6-8 cells without stiffening, while the stiffening
on contact discontinuities in some cases generates numerical
oscillations.
Tangential discontinuities spread over more than 10 cells.

Our tests confirm that slow compound structures with two-dimensional
magnetic field are composed of intermediate shocks (so called ``2-4''
intermediate shocks) followed by slow rarefaction waves.
Finally, tests demonstrate that in two-dimensional magnetohydrodynamics
fast compound structures, which are composed of intermediate shocks
(so called ``1-3'' intermediate shocks) preceded by fast rarefaction
waves, are also possible.

\noindent
{\it subject headings:} hydromagnetics - magnetohydrodynamics:MHD -
methods:numerical - shock waves
\vfill\eject
\section{INTRODUCTION}
Many astronomical objects as diverse as planets, stars, and galaxies all
possess magnetic fields which have important implications on their dynamics
and evolution.
However, in many objects including the Earth the ohmic dissipation time
$t_D=L/\eta^2$ ($\eta$ is the electric resistivity) is smaller than
their ages.
So their magnetic fields should be continuously generated by some dynamo
activity,
otherwise the observed strength cannot be maintained.
The dynamo activity involves patterns of magnetohydrodynamic (MHD) flows
(for example, the interaction of differential rotation and convection as in
some stars or accretion disks) needed to produce spatially coherent magnetic
fields of large scale.
However, such flows are usually highly nonlinear, so with several
exceptions analytic approaches to understand dynamo theory and the origin
of the magnetic field have failed (for review and further references,
see Parker 1979).

With the appearance of fast supercomputers it has become possible to
study the MHD flows numerically, but the development of numerical
techniques to solve MHD equations has been slower due to the intrinsic
complexity of the MHD flows.
For instance, until now most numerical schemes to solve compressible
MHD equations have been based on the methods using artificial viscosity
(\eg DeVore 1991; Lind, Payne \& Meier 1991; Stone \etal~1992; Stone \& Norman
1992)
while schemes to solve compressible hydrodynamic equations have been
based on methods using more sophisticated linear or nonlinear Riemann
solvers (\eg ~Roe 1981; Colella \& Woodward 1984).
It is rather recent that Brio \& Wu (1988; henceforth BW) and
Zachary \& Colella (1992)
developed schemes for MHD equations based on linear Riemann solvers and
Dai \& Woodward (1994; henceforth DW) one based on a nonlinear Riemann solver.
These new schemes have been proved to handle MHD flows better, even
though they are more expensive in CPU time and more difficult to code.

In this paper, we describe a one-dimensional code to solve numerically
the ideal MHD equations.
It is based on an explicit second-order-accurate finite difference scheme
on an Eulerian grid, called the Total Variation Diminishing (TVD) scheme,
which was originally developed for numerical hydrodynamics by Harten (1983).
It utilizes a Roe-type linear Riemann solver described in BW.
Here, we do not attempt to describe the theoretical philosophy of the scheme
which can be found in Harten (1983) and BW.
Instead, we focus on the description of the procedure necessary to write
the code and the test results for its performance.

We also describe a nonlinear Riemann solver for ideal magnetohydrodynamics
(MHDs) which has been developed to provide analytic solutions for shock tube
tests of the code.
It has been built by following a procedure similar to that in DW,
but using initial guesses with the linear eigenvectors
from BW and Zachary \& Colella (1992).

The paper is organized in the following way.
In \S 2 we describe the step by step procedure of the code.
In \S 3 we describe the nonlinear Riemann solver in details.
We intend to make the detailed descriptions of the code and the nonlinear
Riemann solver so others can recover them by following the descriptions.
The tests using shock tubes are presented in \S 4 to demonstrate the ability
of the code to capture discontinuities as well as to follow smooth flows.
The tests also serve to show the performance of the nonlinear Riemann solver.
Finally, discussion is followed in \S 5 including the possible existence
of fast compound structures as well as slow compound structures for
two-dimensional magnetic field structures.
\section{TVD CODE FOR IDEAL MHDS}
%
\subsection{Ideal MHD Equations}
The subject of MHDs describes the dynamics of electrically conducting
fluids in the presence of magnetic fields.
The MHD equations represent coupling of the equations of fluid dynamics
with the Maxwell's equations of electrodynamics.
By neglecting the displacement current, the separation between ions and
electrons, and the effects of electrical resistivity, viscosity, and thermal
conduction, we get the following ideal MHD equations:
$${\partial\rho\over\partial t} + {\bmsy\nabla}\cdot
\left(\rho {\bmit v}\right) = 0, \eqno(2.1)$$
$${\partial{\bmit v}\over\partial t} + {\bmit v}\cdot{\bmsy\nabla}
{\bmit v} +{1\over\rho}{\bmsy \nabla}p - {1\over\rho}
\left({\bmsy\nabla}\times{\bmit B}\right)\times{\bmit B} = 0, \eqno(2.2)$$
$${\partial p\over\partial t} + {\bmit v}\cdot{\bmsy\nabla}p
+ \gamma p{\bmsy\nabla}\cdot{\bmit v} = 0, \eqno(2.3)$$
$${\partial{\bmit B}\over\partial t} - {\bmsy\nabla}\times
\left({\bmit v}\times{\bmit B}\right) = 0, \eqno(2.4)$$
with an additional constraint ${\bmsy\nabla}\cdot{\bmit B}=0$ for the
absence of magnetic monopole (for details, see Shu 1992).
Here, we have chosen units so that factor of $4\pi$ does not appear
in the equations.

For plane-symmetric, or one-dimensional flows exhibiting variation along
the x direction. The equations in (2.1) to (2.4) can be written in conservative
form as
$${\partial{\bmit q}\over\partial t}+{\partial{\bmit F}\over
\partial x}=0, \eqno(2.5)$$
{\textfont1 = \twelvei
       \scriptfont1 = \twelvei \scriptscriptfont1 = \teni
       
$${\bmit q} = \left(\matrix{\rho\cr
                          \rho v_x\cr
                          \rho v_y\cr
                          \rho v_z\cr
                          B_y\cr
                          B_z\cr
                          E\cr}\right)_, \eqno(2.6)$$}
{\textfont1 = \twelvei
       \scriptfont1 = \twelvei \scriptscriptfont1 = \teni
       
$${\bmit F} = \left(\matrix{\rho v_x\cr
                          \rho v_x^2+p^*-B_x^2\cr
                          \rho v_x v_y-B_x B_y\cr
                          \rho v_x v_z-B_x B_z\cr
                          B_y v_x-B_x v_y\cr
                          B_z v_x-B_x v_z\cr
            (E+p^*)v_x-B_x(B_x v_x+B_y v_y+B_z v_z)}\right)_, \eqno(2.7)$$}
where the total pressure and the total energy are given by
$$p^*=p+{1\over2}\left(B_x^2+B_y^2+B_z^2\right), \eqno(2.8)$$
$$E={1\over2}\rho\left(v_x^2+v_y^2+v_z^2\right)+{p\over \gamma-1}
+{1\over2}\left(B_x^2+B_y^2+B_z^2\right). \eqno(2.9)$$
With the state vector, ${\bmit q}$, and the flux function,
${\bmit F}({\bmit q})$, the Jacobian matrix, ${\bmit A}({\bmit q})
=\partial{\bmit F}/\partial{\bmit q}$, is formed.
The above system of equations is called {\it hyperbolic}, since all the
eigenvalues of the Jacobian matrix are real and the corresponding set of
the right eigenvectors is complete (Jeffrey \& Taniuti 1964).
However, the eigenvalues may coincide in some limiting cases (BW).
\subsection{Eigenvalues and Eigenvectors for Plane-Symmetric
MHD Equations}
The first step to build a code based on the Harten's TVD scheme (Harten
1983) for the hyperbolic system of equations in (2.5) is to get the
eigenvalues and the right and left eigenvectors of the Jacobian
${\bmit A}({\bmit q})$.
The seven eigenvalues in nondecreasing order are
$$a_1=v_x-c_f,\eqno(2.10)$$
$$a_2=v_x-c_a,\eqno(2.11)$$
$$a_3=v_x-c_s,\eqno(2.12)$$
$$a_4=v_x,\eqno(2.13)$$
$$a_5=v_x+c_s,\eqno(2.14)$$
$$a_6=v_x+c_a,\eqno(2.15)$$
$$a_7=v_x+c_f,\eqno(2.16)$$
where $c_f$, $c_a$, $c_s$ are the fast, Alfv\'en, and slow characteristic
speeds.
The above represent the seven speeds with which information is propagated
locally by three MHD wave families and an entropy mode.
The three characteristic speeds are expressed as
$$c_a = \sqrt{B_x^2\over\rho}_, \eqno(2.17)$$
$$c_f = \left[{1\over2}\left\{a^2+{B_x^2+B_y^2+B_z^2\over\rho}
+\sqrt{\left(a^2+{B_x^2+B_y^2+B_z^2\over\rho}\right)^2-
4a^2{B_x^2\over\rho}}\right\}\right]^{1\over2}_, \eqno(2.18)$$
$$c_s = \left[{1\over2}\left\{a^2+{B_x^2+B_y^2+B_z^2\over\rho}
-\sqrt{\left(a^2+{B_x^2+B_y^2+B_z^2\over\rho}\right)^2-
4a^2{B_x^2\over\rho}}\right\}\right]^{1\over2}_, \eqno(2.19)$$
where $a$ is the sound speed given by
$$a = \sqrt{\gamma{p\over\rho}}_. \eqno(2.20)$$

The corresponding right eigenvectors are (see, \eg, Jeffrey \& Taniuti 1964)
{\textfont1 = \twelvei
       \scriptfont1 = \twelvei \scriptscriptfont1 = \teni
       
$$R_{v_x\pm c_f} = \left(\matrix{
1\cr
v_x\pm c_f\cr
v_y\mp {B_x B_y c_f\over\rho\left(c_f^2-c_a^2\right)}\cr
v_z\mp {B_x B_z c_f\over\rho\left(c_f^2-c_a^2\right)}\cr
{B_y c_f^2\over\rho\left(c_f^2-c_a^2\right)}\cr
{B_z c_f^2\over\rho\left(c_f^2-c_a^2\right)}\cr
{v_x^2+v_y^2+v_z^2\over2}+h_f^{\pm}\cr
}\right)_, \eqno(2.21)$$}
{\textfont1 = \twelvei
       \scriptfont1 = \twelvei \scriptscriptfont1 = \teni
       
$$R_{v_x\pm c_a} = \left(\matrix{
0\cr
0\cr
\mp B_z~{\rm sign}(B_x)\cr
\pm B_y~{\rm sign}(B_x)\cr
{B_z\over\sqrt{\rho}}\cr
-{B_y\over\sqrt{\rho}}\cr
\mp\left(B_z v_y-B_y v_z\right)~{\rm sign}(B_x)\cr
}\right)_, \eqno(2.22)$$}
{\textfont1 = \twelvei
       \scriptfont1 = \twelvei \scriptscriptfont1 = \teni
       
$$R_{v_x\pm c_s} = \left(\matrix{
1\cr
v_x\pm c_s\cr
v_y\mp {B_x B_y c_s\over\rho\left(c_s^2-c_a^2\right)}\cr
v_z\mp {B_x B_z c_s\over\rho\left(c_s^2-c_a^2\right)}\cr
{B_y c_s^2\over\rho\left(c_s^2-c_a^2\right)}\cr
{B_z c_s^2\over\rho\left(c_s^2-c_a^2\right)}\cr
{v_x^2+v_y^2+v_z^2\over2}+h_s^{\pm}\cr
}\right)_, \eqno(2.23)$$}
{\textfont1 = \twelvei
       \scriptfont1 = \twelvei \scriptscriptfont1 = \teni
       
$$R_{v_x} = \left(\matrix{
1\cr
v_x\cr
v_y\cr
v_z\cr
0\cr
0\cr
{v_x^2+v_y^2+v_z^2\over2}\cr
}\right)_, \eqno(2.24)$$}
where
$$h_f^{\pm}={c_f^2\over\gamma-1}\pm c_f v_x\mp{B_x c_f\left(
B_y v_y+B_z v_z\right)\over\rho\left(c_f^2-c_a^2\right)}+
{\gamma-2\over\gamma-1}\left(c_f^2-a^2\right), \eqno(2.25)$$
$$h_s^{\pm}={c_s^2\over\gamma-1}\pm c_s v_x\mp{B_x c_s\left(
B_y v_y+B_z v_z\right)\over\rho\left(c_s^2-c_a^2\right)}+
{\gamma-2\over\gamma-1}\left(c_s^2-a^2\right). \eqno(2.26)$$

Near the point where either $B_x=0$ or $B_y=B_z=0$, the above set
of the right eigenvectors is not well defined with columns becoming
singular.
By renormalizing the eigenvectors, the singularities can be removed (BW).
The renormalized eigenvectors are
{\textfont1 = \twelvei
       \scriptfont1 = \twelvei \scriptscriptfont1 = \teni
       
$$R_{v_x\pm c_f} = \left(\matrix{
\alpha_f\cr
\alpha_f\left(v_x\pm c_f\right)\cr
\alpha_f v_y\mp\alpha_s\beta_y c_a~{\rm sign}(B_x)\cr
\alpha_f v_z\mp\alpha_s\beta_z c_a~{\rm sign}(B_x)\cr
{\alpha_s\beta_y c_f\over\sqrt{\rho}}\cr
{\alpha_s\beta_z c_f\over\sqrt{\rho}}\cr
\alpha_f{v_x^2+v_y^2+v_z^2\over2}+g_f^{\pm}\cr
}\right)_, \eqno(2.27)$$}
{\textfont1 = \twelvei
       \scriptfont1 = \twelvei \scriptscriptfont1 = \teni
       
$$R_{v_x\pm c_a} = \left(\matrix{
0\cr
0\cr
\mp\beta_z~{\rm sign}(B_x)\cr
\pm\beta_y~{\rm sign}(B_x)\cr
{\beta_z\over\sqrt{\rho}}\cr
-{\beta_y\over\sqrt{\rho}}\cr
\mp\left(\beta_z v_y-\beta_y v_z\right)~{\rm sign}(B_x)\cr
}\right)_, \eqno(2.28)$$}
{\textfont1 = \twelvei
       \scriptfont1 = \twelvei \scriptscriptfont1 = \teni
       
$$R_{v_x\pm c_s} = \left(\matrix{
\alpha_s\cr
\alpha_s\left(v_x\pm c_s\right)\cr
\alpha_s v_y\pm\alpha_f\beta_y a~{\rm sign}(B_x)\cr
\alpha_s v_z\pm\alpha_f\beta_z a~{\rm sign}(B_x)\cr
-{\alpha_f\beta_y a^2\over c_f\sqrt{\rho}}\cr
-{\alpha_f\beta_z a^2\over c_f\sqrt{\rho}}\cr
\alpha_s{v_x^2+v_y^2+v_z^2\over2}+g_s^{\pm}\cr
}\right)_, \eqno(2.29)$$}
{\textfont1 = \twelvei
       \scriptfont1 = \twelvei \scriptscriptfont1 = \teni
       
$$R_{v_x} = \left(\matrix{
1\cr
v_x\cr
v_y\cr
v_z\cr
0\cr
0\cr
{v_x^2+v_y^2+v_z^2\over2}\cr
}\right)_, \eqno(2.30)$$}
where
$$g_f^{\pm}={\alpha_f c_f^2\over\gamma-1}\pm\alpha_f c_f v_x\mp
\alpha_s c_a~{\rm sign}(B_x)\left(\beta_y v_y+\beta_z v_z\right)+
{\gamma-2\over\gamma-1}\alpha_f\left(c_f^2-a^2\right),
\eqno(2.31)$$
$$g_s^{\pm}={\alpha_s c_s^2\over\gamma-1}\pm\alpha_s c_s v_x\pm
\alpha_fa~{\rm sign}(B_x)\left(\beta_y v_y
+\beta_z v_z\right)+{\gamma-2\over\gamma-1}\alpha_s\left(c_s^2
-a^2\right). \eqno(2.32)$$
Here $\alpha$'s and $\beta$'s are given by
$$\alpha_f={\sqrt{c_f^2-c_a^2}\over\sqrt{c_f^2-c_s^2}}_, \eqno(2.33)$$
$$\alpha_s={\sqrt{c_f^2-a^2}\over\sqrt{c_f^2-c_s^2}}_, \eqno(2.34)$$
$$\beta_y = {B_y\over\sqrt{B_y^2+B_z^2}}_, \eqno(2.35)$$
$$\beta_z = {B_z\over\sqrt{B_y^2+B_z^2}}_. \eqno(2.36)$$
At the points where $B_y=B_z=0$, $\beta$'s are defined as the
limiting values, \ie,
$$\beta_y=\beta_z={1\over\sqrt{2}}\qquad{\rm if}\qquad B_y=B_z=0.
\eqno(2.37)$$
Similarly, at the point where $B_y=B_z=0$ and $B_x^2/\rho=a^2$,
$\alpha$'s are defined as
$$\alpha_f=\alpha_s=1\qquad{\rm if}\qquad B_y=B_z=0~~{\rm and}~~
{B_x^2\over\rho}=a^2. \eqno(2.38)$$

Then the left eigenvectors, which are orthonormal to the right
eigenvectors, ${\bmit L}_l\cdot{\bmit R}_m=\delta_{l m}$, are
$$L_{v_x\pm c_f} = \left(l_{v_x\pm c_f}^{(1)},l_{v_x\pm c_f}^{(2)},
l_{v_x\pm c_f}^{(3)},l_{v_x\pm c_f}^{(4)},l_{v_x\pm c_f}^{(5)},
l_{v_x\pm c_f}^{(6)},l_{v_x\pm c_f}^{(7)}\right)_, \eqno(2.39)$$
$$l_{v_x\pm c_f}^{(1)} = {1\over\theta_1}{\alpha_f\over4}a^2v^2
\mp{1\over\theta_2}\left[{\alpha_f\over2}a v_x~{\rm sign}(B_x)
-{\alpha_s\over2}c_s\left(\beta_y v_y+\beta_z v_z\right)\right]_,
\eqno(2.40)$$
$$l_{v_x\pm c_f}^{(2)} = -{1\over\theta_1}{\alpha_f\over2}a^2v_x
\pm{1\over\theta_2}{\alpha_f\over2}a~{\rm sign}(B_x), \eqno(2.41)$$
$$l_{v_x\pm c_f}^{(3)} = -{1\over\theta_1}{\alpha_f\over2}a^2v_y
\mp{1\over\theta_2}{\alpha_s\over2}\beta_yc_s, \eqno(2.42)$$
$$l_{v_x\pm c_f}^{(4)} = -{1\over\theta_1}{\alpha_f\over2}a^2v_z
\mp{1\over\theta_2}{\alpha_s\over2}\beta_zc_s, \eqno(2.43)$$
$$l_{v_x\pm c_f}^{(5)} = {1\over\theta_1}{\alpha_s\over2}\beta_yc_f
\left(c_s^2+{2-\gamma\over\gamma-1}a^2\right)\sqrt{\rho}, \eqno(2.44)$$
$$l_{v_x\pm c_f}^{(6)} = {1\over\theta_1}{\alpha_s\over2}\beta_zc_f
\left(c_s^2+{2-\gamma\over\gamma-1}a^2\right)\sqrt{\rho}, \eqno(2.45)$$
$$l_{v_x\pm c_f}^{(7)} = {1\over\theta_1}{\alpha_f\over2}a^2, \eqno(2.46)$$
$$L_{v_x\pm c_a} = \left(l_{v_x\pm c_a}^{(1)},l_{v_x\pm c_a}^{(2)},
l_{v_x\pm c_a}^{(3)},l_{v_x\pm c_a}^{(4)},l_{v_x\pm c_a}^{(5)},
l_{v_x\pm c_a}^{(6)},l_{v_x\pm c_a}^{(7)}\right)_, \eqno(2.47)$$
$$l_{v_x\pm c_a}^{(1)} = \pm{\beta_zv_y-\beta_yv_z\over2}{\rm sign}(B_x),
\eqno(2.48)$$
$$l_{v_x\pm c_a}^{(2)} = 0, \eqno(2.49)$$
$$l_{v_x\pm c_a}^{(3)} = \mp{\beta_z\over2}{\rm sign}(B_x), \eqno(2.50)$$
$$l_{v_x\pm c_a}^{(4)} = \pm{\beta_y\over2}{\rm sign}(B_x), \eqno(2.51)$$
$$l_{v_x\pm c_a}^{(5)} = {\beta_z\sqrt{\rho}\over2}, \eqno(2.52)$$
$$l_{v_x\pm c_a}^{(6)} = -{\beta_y\sqrt{\rho}\over2}, \eqno(2.53)$$
$$l_{v_x\pm c_a}^{(7)} = 0, \eqno(2.54)$$
$$L_{v_x\pm c_s} = \left(l_{v_x\pm c_s}^{(1)},l_{v_x\pm c_s}^{(2)},
l_{v_x\pm c_s}^{(3)},l_{v_x\pm c_s}^{(4)},l_{v_x\pm c_s}^{(5)},
l_{v_x\pm c_s}^{(6)},l_{v_x\pm c_s}^{(7)}\right)_, \eqno(2.55)$$
$$l_{v_x\pm c_s}^{(1)} = {1\over\theta_1}{\alpha_s\over4}c_f^2v^2
\mp{1\over\theta_2}\left[{\alpha_s\over2}c_a v_x~{\rm sign}(B_x)
+{\alpha_f\over2}c_f(\beta_yv_y+\beta_zv_z)\right]_, \eqno(2.56)$$
$$l_{v_x\pm c_s}^{(2)} = -{1\over\theta_1}{\alpha_s\over2}c_f^2v_x
\pm{1\over\theta_2}{\alpha_s\over2}c_a~{\rm sign}(B_x), \eqno(2.57)$$
$$l_{v_x\pm c_s}^{(3)} = -{1\over\theta_1}{\alpha_s\over2}c_f^2v_y
\pm{1\over\theta_2}{\alpha_f\over2}\beta_yc_f, \eqno(2.58)$$
$$l_{v_x\pm c_s}^{(4)} = -{1\over\theta_1}{\alpha_s\over2}c_f^2v_z
\pm{1\over\theta_2}{\alpha_f\over2}\beta_zc_f, \eqno(2.59)$$
$$l_{v_x\pm c_s}^{(5)} = -{1\over\theta_1}{\alpha_f\over2}\beta_yc_f
\left(c_f^2+{2-\gamma\over\gamma-1}a^2\right)\sqrt{\rho}, \eqno(2.60)$$
$$l_{v_x\pm c_s}^{(6)} = -{1\over\theta_1}{\alpha_f\over2}\beta_zc_f
\left(c_f^2+{2-\gamma\over\gamma-1}a^2\right)\sqrt{\rho}, \eqno(2.61)$$
$$l_{v_x\pm c_s}^{(7)} = {1\over\theta_1}{\alpha_s\over2}c_f^2, \eqno(2.62)$$
$$L_{v_x} = \left(l_{v_x}^{(1)},l_{v_x}^{(2)},l_{v_x}^{(3)},
l_{v_x}^{(4)},l_{v_x}^{(5)},l_{v_x}^{(6)},l_{v_x}^{(7)}\right)_, \eqno(2.63)$$
$$l_{v_x}^{(1)} = 1-{1\over\theta_1}{\alpha_f^2a^2+\alpha_s^2c_f^2\over2}
v^2, \eqno(2.64)$$
$$l_{v_x}^{(2)} = {1\over\theta_1}\left(\alpha_f^2a^2+\alpha_s^2c_f^2\right)
v_x, \eqno(2.65)$$
$$l_{v_x}^{(3)} = {1\over\theta_1}\left(\alpha_f^2a^2+\alpha_s^2c_f^2\right)
v_y, \eqno(2.66)$$
$$l_{v_x}^{(4)} = {1\over\theta_1}\left(\alpha_f^2a^2+\alpha_s^2c_f^2\right)
v_z, \eqno(2.67)$$
$$l_{v_x}^{(5)} = {1\over\theta_1}\alpha_f\alpha_s\beta_yc_f
\left(c_f^2-c_s^2\right)\sqrt{\rho}, \eqno(2.68)$$
$$l_{v_x}^{(6)} = {1\over\theta_1}\alpha_f\alpha_s\beta_zc_f
\left(c_f^2-c_s^2\right)\sqrt{\rho}, \eqno(2.69)$$
$$l_{v_x}^{(7)} = -{1\over\theta_1}\left(\alpha_f^2a^2+\alpha_s^2c_f^2\right),
\eqno(2.70)$$
where
$$\theta_1 = \alpha_f^2a^2\left(c_f^2+{2-\gamma\over\gamma-1}a^2\right)
+\alpha_s^2c_f^2\left(c_s^2+{2-\gamma\over\gamma-1}a^2\right)_,
\eqno(2.71)$$
$$\theta_2 = \alpha_f^2c_fa~{\rm sign}(B_x)+\alpha_s^2c_sc_a~
{\rm sign}(B_x), \eqno(2.72)$$
$$v^2 = v_x^2+v_y^2+v_z^2. \eqno(2.73)$$

Note that, with the above normalization, some columns are not continuous.
In order to force them to be continuous, the following term
$${\rm sign}(B_T) = \cases{ 1, &if~~~$B_y>0$~~~or~~~$B_y=0$ and $B_z>0$\cr
                           -1, &if~~~$B_y<0$~~~or~~~$B_y=0$ and $B_z<0$\cr}_,
\eqno(2.74)$$
is multiplied $R_{v_x\pm c_s}$ and $L_{v_x\pm c_s}$ if $a^2>c_a^2$ and
to $R_{v_x\pm c_f}$ and $L_{v_x\pm c_f}$ if $a^2<c_a^2$.

It is interesting to see how the eigenvectors for the MHD equations
reduce into those for the hydrodynamic equations in the limit
$B_y\rightarrow0$ and $B_z\rightarrow0$.
If $a^2>c_a^2$, $R_{v_x\pm c_f}\rightarrow R^{\rm HD}_{v_x\pm a}$
and $R_{v_x}\rightarrow R^{\rm HD}_{v_x}$, that is, the characteristics
associated with MHD fast waves become those associated with hydrodynamic
sound waves.
The other two eigenvectors for the three-dimensional hydrodynamic equations
(Roe 1981) are obtained from the following combinations of those for
slow and Alfv\'en waves,
{\textfont1 = \twelvei
       \scriptfont1 = \twelvei \scriptscriptfont1 = \teni
       
$${\sqrt{2}~{\rm sign}(B_x)\over4}\left[{R_{v_x+c_s}-R_{v_x-c_s}\over a}
-\left(R_{v_x+c_a}-R_{v_x-c_a}\right)\right] = \left(\matrix{
0\cr 0\cr 1\cr 0\cr 0\cr 0\cr v_y\cr}\right)_, \eqno(2.75)$$}
{\textfont1 = \twelvei
       \scriptfont1 = \twelvei \scriptscriptfont1 = \teni
       
$${\sqrt{2}~{\rm sign}(B_x)\over4}\left[{R_{v_x+c_s}-R_{v_x-c_s}\over a}
+\left(R_{v_x+c_a}-R_{v_x-c_a}\right)\right] = \left(\matrix{
0\cr 0\cr 0\cr 1\cr 0\cr 0\cr v_z\cr}\right)_. \eqno(2.76)$$}
Similarly, if $a^2<c_a^2$, $R_{v_x\pm c_s}\rightarrow R^{\rm HD}_{v_x\pm a}$
and $R_{v_x}\rightarrow R^{\rm HD}_{v_x}$, that is, the characteristics
associated with MHD slow waves become those associated with hydrodynamic
sound waves.
The other two eigenvectors for the three-dimensional hydrodynamic equations
are obtained from the following combinations of those for fast and Alfv\'en
waves,
{\textfont1 = \twelvei
       \scriptfont1 = \twelvei \scriptscriptfont1 = \teni
       
$${\sqrt{2}~{\rm sign}(B_x)\over4}\left[-{R_{v_x+c_f}-R_{v_x-c_f}\over c_a}
-\left(R_{v_x+c_a}-R_{v_x-c_a}\right)\right] = \left(\matrix{
0\cr 0\cr 1\cr 0\cr 0\cr 0\cr v_y\cr}\right)_, \eqno(2.77)$$}
{\textfont1 = \twelvei
       \scriptfont1 = \twelvei \scriptscriptfont1 = \teni
       
$${\sqrt{2}~{\rm sign}(B_x)\over4}\left[-{R_{v_x+c_f}-R_{v_x-c_f}\over c_a}
+\left(R_{v_x+c_a}-R_{v_x-c_a}\right)\right] = \left(\matrix{
0\cr 0\cr 0\cr 1\cr 0\cr 0\cr v_z\cr}\right)_. \eqno(2.78)$$}

In the cases with purely two-dimensional magnetic fields and motions
(\ie, $B_z=v_z=0$), the eigenvectors associated with Alfv\'en waves becomes
{\textfont1 = \twelvei
       \scriptfont1 = \twelvei \scriptscriptfont1 = \teni
       
$$R_{v_x\pm c_a} = \left(\matrix{0\cr 0\cr 0\cr \pm{\rm sign}(B_x)\cr
0\cr -{1\over\sqrt{\rho}}\cr 0\cr}\right)_. \eqno(2.79)$$}
By combining them properly, they become $(0,0,0,1,0,0,0)^T$ and
$(0,0,0,0,0,1,0)^T$ which are trivial.
This indicates that the flows with two-dimensional magnetic fields
do not produce structures associated with Alfv\'en modes
like rotational discontinuities (see \S 4 and 5 for more discussions)
\subsection{A Second-Order Explicit TVD Scheme}
Here, we describe briefly the procedure to build the MHD-TVD code
with the eigenvalues and eigenvectors in the previous subsection.
The purpose of this section is to provide a short but complete
description of steps needed to build a code by the TVD scheme.
For the details, \eg, why and how each step works, the choices
for the values of internal parameters, etc., refer to the original
work (Harten 1983).

The convention for indices used in this subsection is the following.
The superscript $n$ represents the time step.
The subscript $i$ indicates the quantities defined in the cell centers
while $i+1/2$ identifies those defined on the cell boundaries.
The subscript $k$ represents the characteristic fields with the order
that $k=1$ is for the field associated with the eigenvalue $v_x-c_f$,
$k=2$ for the field with $v_x-c_a$, $k=3$ for the field with $v_x-c_s$,
$k=4$ for the field with $v_x$, $k=5$ for the field with $v_x+c_s$, $k=6$
for the field with $v_x+c_a$, and $k=7$ for the field with $v_x+c_f$.

In a code based on the TVD scheme, the physical quantities are defined
in the cell centers while the fluxes are computed on the cell boundaries.
Implementation of Roe's linearization technique would result in a
particular form of the averaged physical quantities in the cell
boundaries (Roe 1981).
However, as pointed by BW, it is not possible to
derive the analytic form of the averaged quantities in MHDs for general
cases with the adiabatic index $\gamma\ne2$.
Instead, we should modify the Roe's scheme by using the following
simple averaging scheme,
$$\rho_{i+{1\over2}} = {\rho_i+\rho_{i+1} \over 2}, \eqno(2.80)$$
$$v_{x,i+{1\over2}} = {v_{x,i}+v_{x,i+1} \over 2}, \eqno(2.81)$$
$$v_{y,i+{1\over2}} = {v_{y,i}+v_{y,i+1} \over 2}, \eqno(2.82)$$
$$v_{z,i+{1\over2}} = {v_{z,i}+v_{z,i+1} \over 2}, \eqno(2.83)$$
$$B_{y,i+{1\over2}} = {B_{y,i}+B_{y,i+1} \over 2}, \eqno(2.84)$$
$$B_{z,i+{1\over2}} = {B_{z,i}+B_{z,i+1} \over 2}, \eqno(2.85)$$
$$p^*_{i+{1\over2}} = {p^*_i+p^*_{i+1} \over 2}. \eqno(2.86)$$
Then, other quantities like momentum, gas pressure, total energy, etc
are calculated by combining the above quantities.
Our tests for the cases with $\gamma=2$ indicated that the above
simple averaging would do just as well when compared to the full
implementation of Roe's linearization technique.

The state vector $\bmit q$ at the cell center is updated by calculating
the modified flux $\bar{\bmit f}$ at the cell boundaries as follows:
$${\bmit q}_i^{n+1}={\bmit q}_i^n - {\Delta t^n\over \Delta x}
(\bar{\bmit f}_{i+{1\over2}}-\bar{\bmit f}_{i-{1\over2}}), \eqno(2.87)$$
$$\bar{\bmit f}_{i+{1\over2}} = {1\over2}\left[{\bmit F}({\bmit q}_i^n)
+{\bmit F}({\bmit q}_{i+1}^n)\right]-{\Delta x\over2\Delta t^n}\sum_{k=1}^7
\beta_{k,i+{1\over2}}{\bmit R}_{k,i+{1\over2}}^n, \eqno(2.88)$$
$$\beta_{k,i+{1\over2}} = Q_k\left({\Delta t^n\over\Delta x}
a_{k,i+{1\over2}}^n+\gamma_{k,i+{1\over2}}\right)\alpha_{k,i+{1\over2}}
-(g_{k,i}+g_{k,i+1}), \eqno(2.89)$$
$$\alpha_{k,i+{1\over2}} = {\bmit L}_{k,i+{1\over2}}^n\cdot
({\bmit q}_{i+1}^n-{\bmit q}_i^n), \eqno(2.90)$$
{\textfont1 = \twelvei
       \scriptfont1 = \twelvei \scriptscriptfont1 = \teni
       
$$\gamma_{k,i+{1\over2}} = \cases{{g_{k,i+1}-g_{k,i}\over
\alpha_{k,i+{1\over2}}}_,&for $\alpha_{k,i+{1\over2}}\neq0$\cr
0, &for $\alpha_{k,i+{1\over2}}=0$\cr}_, \eqno(2.91)$$}
$$g_{k,i} = {\rm sign}({\tilde g}_{k,i+{1\over2}})~
{\rm max}\left[0,~{\rm min}\left\{|{\tilde g}_{k,i+{1\over2}}|,~
{\tilde g}_{k,i-{1\over2}}{\rm sign}({\tilde g}_{k,i+{1\over2}})
\right\}\right], \eqno(2.92)$$
$${\tilde g}_{k,i+{1\over2}} = {1\over2}
\left[Q_k({\Delta t^n\over\Delta x}a_{k,i+{1\over2}}^n)
-({\Delta t^n\over\Delta x}a_{k,i+{1\over2}}^n)^2\right]
\alpha_{k,i+{1\over2}}, \eqno(2.93)$$
{\textfont1 = \twelvei
       \scriptfont1 = \twelvei \scriptscriptfont1 = \teni
       
$$Q_k(\chi) = \cases{{\chi^2\over4\varepsilon}+\varepsilon,
& for $|\chi|<2\varepsilon$\cr
|\chi|, & for $|\chi|\geq2\varepsilon$\cr}_, \eqno(2.94)$$}
$$\varepsilon = \cases{0.1, & for $k=1$ and $7$ \cr
                       0.2,  & for $k=2$ and $6$ \cr
                       0.1, & for $k=3$ and $5$ \cr
                       0.0   & for $k=4$\cr}_. \eqno(2.95)$$
Here, the time step $\Delta t^n$ is restricted by the usual Courant
condition for the stability, $\Delta t^n=C_{\rm cour}\Delta x
/{\rm Max}(|v^n_{x,i+{1\over2}}|+c^n_{f,i+{1\over2}})$ with $C_{\rm cour}<1$.
Typically we use $C_{\rm cour}=0.8$.

The rotational discontinuities represented by the $k=2$ and $6$ fields are
steepened by replacing $g_{k,i}$ with $g_{k,i}+\theta_{k,i}\bar g_{k,i}$:
{\textfont1 = \twelvei
       \scriptfont1 = \twelvei \scriptscriptfont1 = \teni
       
$$\theta_{k,i} = \cases{{|\alpha_{k,i+{1\over2}}-\alpha_{k,i-{1\over2}}|
\over|\alpha_{k,i+{1\over2}}|+|\alpha_{k,i-{1\over2}}|}_,
& for $(|\alpha_{k,i+{1\over2}}|+|\alpha_{k,i-{1\over2}}|)\neq0$\cr
0, & for $(|\alpha_{k,i+{1\over2}}|+|\alpha_{k,i-{1\over2}}|)=0$\cr}_,
\eqno(2.96)$$}
$$\bar g_{k,i} = {\rm sign}(\alpha_{k,i+{1\over2}})~
{\rm max}\left[0,~{\rm min}\left\{{\rm sign}(\alpha_{k,i+{1\over2}})
\sigma_{k,i-{1\over2}}\alpha_{k,i-{1\over2}},~
\sigma_{k,i+{1\over2}}|\alpha_{k,i+{1\over2}}|\right\}\right],
\eqno(2.97)$$
$$\sigma_{k,i+{1\over2}} = {1\over2}\left[1-\left|{\Delta t^n\over\Delta x}
a_{k,i+{1\over2}}\right|\right]_. \eqno(2.98)$$
With the above steepening the rotational discontinuities are resolved
within 3-5 cells, otherwise within 6-8 cells.
The contact discontinuities represented by the $k=4$ field could be also
steepened by a similar scheme.
However, our tests showed that the price for steepening the contact
discontinuities is additional numerical oscillations.
\section{A NONLINEAR MHD RIEMANN SOLVER}
As a means to test quantitatively the MHD-TVD code and to understand
more fully the properties of ideal MHD flows along one-dimension,
we developed an accurate nonlinear MHD Riemann solver.
It is similar in many respects to that described in DW.
The most important improvement in our MHD Riemann solver is that it treats
fast and slow rarefactions properly, instead of approximating them
as ``rarefaction shocks'' as did the solver presented in DW.
We present here only essential details of the ``solver'', referring
readers to DW for other relevant background.

As with hydrodynamic Riemann solvers, the construction of an MHD Riemann
solver is based on the idea that two adjacent arbitrary states will
evolve into a set of uniform states separated by left and right facing
shocks and rarefactions.
For the MHD problems, there are a total of eight states including the
original pair.
They are separated by six structures representing left and right
propagating shocks or rarefactions of the three wave families and a
structure representing the entropy mode (or the characteristic field
associated with the eigenvalues $v_x$ in the discussion of the previous
section).
The initial boundary moves with the structure of the entropy mode
which becomes a contact discontinuity or, in degenerate cases, a tangential
discontinuity (see \S 4 for more discussion).
In fact the eigenstates developed in \S 2 provide approximate solutions
to this problem.
The particular difficulty in the MHD Riemann problem is that the equations
are not strictly hyperbolic nor strictly convex (see the discussion in BW).
In practice this means that the wave speeds of two families may sometimes
coincide, and that compound wave structures involving both shocks
and rarefactions may sometimes develop.
For the most part we can ignore this aspect, and consider those
situations as special cases.
Otherwise we can approach the Riemann problem in pretty much the same
manner as for the hydrodynamical problem, except for the larger number of
waves to consider.

The solution of the problem is obtained by finding the required set of
fast and slow shock jumps and rarefactions together with rotational
discontinuities that self-consistently lead to a proper contact
discontinuity or tangential discontinuity at the ``center'' of the
structure.
Our procedure consists of taking an initial guess for the resolved
states of the (six interior) zones, and then iterating towards the proper jump
conditions of the contact discontinuity or the tangential discontinuity.
For this discussion, the left initial state is identified with zone 1 and
the right initial state is identified with zone 8.
Fast structures separate zones 1 and 2 as well as zones 7 and 8.
Rotational discontinuities separate zones 2 and 3 as well as zones 6 and 7.
Zones 3 and 4 and zones 5 and 6 are each separated by slow structures.
The contact discontinuity or the tangential discontinuity demarcates
zones 4 and 5.

Discontinuities are easier to handle than rarefactions, so we outline
the methods to compute discontinuities first.
The problem is conceptually simpler in Lagrangian mass coordinates,
where the jump conditions across a discontinuity are
(\eg, equation (2.7), DW):
$$W[V] = - [v_x],\eqno(3.1)$$
$$W[v_x] = [p^* - B_x^2],\eqno(3.2)$$
$$W[v_y] = - B_x [B_y],\eqno(3.3)$$
$$W[v_z] = - B_x [B_z],\eqno(3.4)$$
$$W[VB_y] = - B_x [v_y],\eqno(3.5)$$
$$W[VB_z] = - B_x [v_z],\eqno(3.6)$$
$$W[VE] = [v_x p^*] - B_x[B_x v_x + B_y v_y + B_z v_z],\eqno(3.7)$$
where $[Q] = Q_d - Q_u$ is the difference between the downstream
and upstream values of a quantity $Q$, $W = - (\rho  v_x)_u
= - (\rho v_x)_d $ is the Lagrangian speed of the discontinuity in the
mass coordinates, and $V = 1/\rho$.
The other quantities are as defined in \S 2.
Thus, given the upstream state and an estimate of the Lagrangian speed of
the discontinuity, the downstream state can be computed.

The fast shock speed $W_f$ and the slow shock speed $W_s$ in the
Lagrangian mass coordinates are given in DW as
$$W_{f,s}^2 = {1\over 2} {1\over {1 + S_0}} \left\lbrace { (C_s^2 + C_f^2 +
S_1) \pm \sqrt{(C_s^2 + C_f^2 + S_1)^2 - 4 (1 + S_0)(C_s^2 C_f^2 - S_2)} }
\right\rbrace, \eqno(3.8)$$
where $C_{f} = \rho c_{f}$ and $C_{s} = \rho c_{s}$.
The upper (lower) sign refers to fast (slow) waves.
The coefficients $S_0, S_1, S_2$ can be written in terms of the jump in
tangential magnetic field across the shock, $[B_\perp]$, as
$$S_0 = - {1\over 2} (\gamma - 1) {{[B_\perp]}\over {B_\perp}},\eqno(3.9)$$
$$S_1 = {1\over 2}\left\lbrace -(\gamma - 2)C_\perp^2 {[B_\perp]\over B_\perp}
+2 C_o^2 - (\gamma - 4)C_\perp^2 - 2\gamma C_a^2 \right\rbrace {{[B_\perp]}
\over {B_\perp}},\eqno(3.10)$$
$$S_2 = {1\over 2}\left\lbrace {C_a^2 ([B_\perp])^2\over V} + {(\gamma+2)
C_\perp C_a^2 [B_\perp]\over\sqrt{V}} + C_\perp^2 C_a^2 (\gamma+1)
+(\gamma + 1) C_a^4 - 2 C_o^2 C_a^2 \right\rbrace {{[B_\perp]}\over {B_\perp}},
\eqno(3.11)$$
where $C_o = \rho a$ is the Lagrangian sound speed, $C_a = \rho c_a$ is
the Lagrangian Alfv\'en speed, $C_\perp = \sqrt{\rho} B_\perp$,
and $B_\perp = \sqrt{B_y^2 + B_z^2}$.
All quantities except $[B_\perp]$ in equations (3.9) to (3.11) are
referred to the state upstream of the shock.  The expressions used in
equations (3.9) to (3.11) are equivalent to those given by DW, but are
somewhat simpler and in our
experience have provided more robust behavior in the Riemann solver,
particularly in switch-on or switch-off shocks.
An expression analogous to equation (3.8) appropriate to
the special case $B_x = 0$ (magnetosonic shocks) is also given by DW.
As is well known, fast and slow shocks do not alter the plane of the magnetic
field.

Rotational discontinuities, which are not compressive, can be handled by
setting the jump $[v_x] = 0$ in equations (3.1) and (3.2).
That leads necessarily to $W_a = \pm C_a$.
The jump conditions required at the contact discontinuity can be found
by setting $W = 0$ in equations (3.1) to (3.7).
With these results, it is clear that if the jumps of $[B_\perp]$ across
shocks and the rotations of $[B_\perp]$ across rotational discontinuities
are known then it should be possible to exactly determine the structure of
any ideal MHD Riemann problem that involves only discontinuous interfaces.

DW included rarefactions in their MHD Riemann solver by assuming they could
also be treated as discontinuities ({\it i.e.,} as ``rarefaction shocks'').
So long as the rarefactions are weak this is reasonably accurate, but
not exact.
In fact, it is also straightforward to include fast and slow rarefactions
exactly, just as for hydrodynamics.
By conserving all Riemann invariants through the rarefactions except
that associated with the particular wave involved, one can derive a simple
set of differential equations to be integrated through the rarefactions.
The transitions computed in this way then replace the jumps given in
equations (3.1) to (3.7).

The relations appropriate to right facing (upper sign) and
left facing (lower sign) fast rarefactions are (\eg, Jeffrey 1966)
$$C^{\prime}_o = -{{\gamma + 1}\over{2}}\sqrt{\rho} {{C_\perp C^2_s}\over
{C_o ( C^2_s~-~C^2_a)}} = {{\gamma + 1}\over{2}}\sqrt{\rho}
{{C^2_s (C^2_f~-~C^2_a)}\over{C^2_a C_\perp C_o}} = {{\gamma + 1}\over{2}}
{{\rho p^{\prime}}\over{C_o}},\eqno(3.12)$$
$$u^{\prime}_x = \mp {{1}\over{\sqrt{\rho}}} {{C_\perp C^2_a}\over {C_f
(C^2_s~-~C^2_a)}}
= \pm {{C^2_f~-~C^2_a}\over{\sqrt{\rho} C_\perp C_f}}
= \pm {{2}\over{\gamma + 1}}{{C_o C^2_a}\over{C^2_s C_f}}{{C^{\prime}_o}
\over{\rho}},\eqno(3.13)$$
$${{u^{\prime}_y}\over{\cos{\psi}}} = {{u^{\prime}_z}\over{\sin{\psi}}}
= \mp {{1}\over{\sqrt{\rho}}} {{C_a}\over{C_f}}.\eqno(3.14)$$
For right and left facing slow rarefactions one finds
$$C^{\prime}_o = -{{\gamma + 1}\over{2}}\sqrt{\rho} {{C_\perp C^2_f}\over
{C_o ( C^2_f~-~C^2_a)}} = {{\gamma + 1}\over{2}}\sqrt{\rho}
{{C^2_f (C^2_s~-~C^2_a)}\over{C^2_a C_\perp C_o}} = {{\gamma + 1}\over{2}}
{{\rho p^{\prime}}\over{C_o}},\eqno(3.15)$$
$$u^{\prime}_x = \mp {{1}\over{\sqrt{\rho}}} {{C_\perp C^2_a}\over {C_s
(C^2_f~-~C^2_a)}}
= \pm {{C^2_s~-~C^2_a}\over{\sqrt{\rho} C_\perp C_s}}
= \pm {{2}\over{\gamma + 1}}{{C_o C^2_a}\over{C^2_f C_s}}{{C^{\prime}_o}
\over{\rho}},\eqno(3.16)$$
$${{u^{\prime}_y}\over{\cos{\psi}}} = {{u^{\prime}_z}\over{\sin{\psi}}}
= \mp {{1}\over{\sqrt{\rho}}} {{C_a}\over{C_s}}.\eqno(3.17)$$
In these expressions, $\tan{\psi} = B_z/B_y$, and primes represent
derivatives with respect to $B_\perp$.
It is also useful to have the relationship
$$(C^2_f~-~C^2_a)(C^2_s~-~C^2_a) = - C^2_a C^2_\perp.\eqno(3.18)$$

{\parskip=4pt
Our procedure for obtaining an accurate solution to a Riemann problem
is similar to that employed by DW.
We utilize the following steps:\hfill\break

\noindent 1) As an initial guess, we found it convenient and mostly reliable to
use
the eigenvectors defined in \S 2.
In particular if the state vector, ${\bmit q}$, defined in equation (2.6) is
represented in each of the intermediate regions by ${\bmit q}(i)$
$(i = 2,3,...7)$ then an estimate of ${\bmit q}(i)$ is simply
$${\bmit q}(i) = {\bmit q}(8) - \sum_{k=i+1}^7 \alpha_k {\bmit R}_k,
\eqno(3.19)$$
where $\alpha_k$ is defined as in equation (2.90), and ${\bmit q}(8)$ is
the initial right state.
Occasionally the states found by equation (3.19) can have unphysical
properties ({\it i.e.}, jumps in $B_\perp$ that violate equations (3.1)
to (3.7)), so some simple physical constraints need to be applied.\hfill\break

\noindent 2) The quantities $B_\perp(2)$, $B_\perp(4)$, $B_\perp(7)$ and
$\psi(3)$
from that solution are applied to determine jumps across each of the six
waves (\eg~$[B_\perp]_{1\rightarrow 2}$), subject to the constraints that
$B_\perp(5)=B_\perp(4)$ and $\psi(4-6) = \psi(3)$.
This is accomplished by starting with the two fast waves, followed by the
two rotations of $B_\perp$ and finally the two slow waves.\hfill\break

\noindent 3) For the solution to be considered exact, we require that the
resulting
jump conditions at the contact discontinuity precisely satisfy those
expected from equations (3.1) to (3.7).
In particular, we test if all the jumps $[v_x]$, $[v_y]$, $[v_z]$ and
$[p^*] = 0$.\hfill\break

\noindent 4) If the contact discontinuity jump conditions are not satisfied,
we vary $B_\perp(2)$, $B_\perp(4)$, $B_\perp(7)$ and $\psi(3)$ using a
Newton-Raphson scheme based on a numerical approximation to the associated
Jacobian matrix for an improved guess in the quantities $B_\perp(2)$,
$B_\perp(4)$,
$B_\perp(7)$ and $\psi(3)$.\hfill\break

\noindent 5) The procedure, beginning with step 2), is then repeated until
convergence is obtained.
The accuracy of the initial guess is the single most important aspect
controlling the number of iterations required.
Once a reasonably approximate solution is found, usually only a couple of
iterations lead to very good convergence.\hfill\break}

When $B_x = 0$ the same scheme is applied, but in somewhat simplified form,
since there are no
slow wave features or rotational discontinuities and the constraint
$\psi(4) = \psi(5)$ is removed.

With these procedures, we are able in most cases to obtain solutions such
that the contact discontinuity jump conditions are very well satisfied,
even to near the limits of machine accuracy.
In practice, we establish a convergence criterion of $10^{-6}$ relative
to the larger of a zone averaged $v_x$ or fast wave speed for velocities
or total pressure for $[p^*]$.
Only for Riemann problems involving switch-off or switch-on waves, we
are significantly limited, since an exact switch-on or switch-off feature
requires either the upstream or downstream flow speed to exactly equal the
Alfv\'en speed.
In those cases, even very small errors in $[B_\perp]$ lead to significant
errors in $W$,
and so the remaining downstream state variables.
We found that we could obtain precise solutions only if we permitted the
smaller $B_\perp$ next to such a wave to have a value $\sim 10^{-3}$ of
the larger $B_\perp$.
These waves are for almost any practical purpose indistinguishable from exact
switch-waves, however.

We will discuss tests of the Riemann solver along with the tests of
the MHD TVD code in the next section, since we test them against each other
in many cases.
Suffice it to say here that, to start with, we examined our Riemann solver
against all those solutions presented in DW and against a number of other
examples kindly given to us by Dr.~Dai which were also generated with the
Riemann solver described in DW.
In the cases involving only discontinuities, our solutions agree exactly
with the DW results.
When there are rarefactions, we find some differences.
But those are attributable to the fact that we treated rarefactions exactly,
whereas they did not.
Mostly those differences are relatively small, indicating that treating
rarefactions as shocks would produce a reasonable result in an approximate
MHD Riemann solver.
\section{NUMERICAL TESTS}
To test the code described in \S 2 as well as the Riemann solver described
in \S 3, we chose MHD shock tube problems including those considered in BW
and DW.
In all the tests, we set the adiabatic index $\gamma=5/3$ and used
a one-dimensional box with $x=[0,1]$.
The numerical calculations were done with a Courant constant 0.8
and without the stiffening of the contact discontinuity.
The results of the numerical calculations with the code are plotted as dots
and the analytic solutions of the Riemann solver are plotted as lines.
The plotted quantities are density, gas pressure, total (thermal, kinetic,
and magnetic) energy, $x$-velocity (parallel to the direction of structure
propagation), $y$-velocity, $z$-velocity, $y$-magnetic field, $z$-magnetic
field, and the orientation angle of tangential magnetic field
$(\psi=\tan^{-1}(B_z/B_y))$
in the plane perpendicular to the propagation vector.
Numerical values of the analytic solutions in the regions between the
structures (\eg ~between the left moving fast shock and the left moving
rotational discontinuity, etc) are listed in the tables with the same labels as
the figures.  To simplify the discussion of this section we refer to solutions
as {\it two-dimensional} when the magnetic field remains in one plane through
the
entire structure, or {\it three-dimensional} when the field cannot be so
described.

The first set of tests (also found in DW) has been done with two-dimensional
field and
velocity structure in the $x-y$ plane but without change in the direction
of the tangential magnetic field ($B_y$ in these tests).
Fig.~1a shows the solution of the MHD shock tube test with the left state
($\rho$, $v_x$, $v_y$, $v_z$, $B_y$, $B_z$, $E$) = ($1$, $10$, $0$, $0$,
$5/\sqrt{4\pi}$, $0$, $20$) and the right state ($1$, $-10$, $0$, $0$,
$5/\sqrt{4\pi}$, $0$, $1$) with $B_x=5/\sqrt{4\pi}$ at time $t=0.08$.
The plot shows a pair of fast shocks, a left facing slow rarefaction, a right
facing slow shock and a contact discontinuity.
Fig.1~b shows the solution of the MHD shock tube test with the left state
($\rho$, $v_x$, $v_y$, $v_z$, $B_y$, $B_z$, $E$) = ($1$, $0$, $0$, $0$,
$5/\sqrt{4\pi}$, $0$, $1$) and the right state ($0.1$, $0$, $0$, $0$,
$2/\sqrt{4\pi}$, $0$, $10$) with $B_x=3/\sqrt{4\pi}$ at time $t=0.03$.
The plot shows one fast shock and one fast rarefaction, one slow shock
and one slow rarefaction, and a contact discontinuity.
As expected from the two-dimensional nature of the field and velocity
structure, there is no rotational discontinuity in either of these flows.
In the numerical calculations, fast shocks that are strong with a large
parallel velocity jump, $[v_x]$, are resolved within 2-4 cells, while
slow shocks that are weak with a small jump require more cells to be resolved.
Contact discontinuity spread typically spread over 6-8 or so cells.

The second set of tests (also from DW) involves three-dimensional field and
velocity
structure where the magnetic field plane rotates.
The solution of the MHD shock tube test with the left state
($\rho$, $v_x$, $v_y$, $v_z$, $B_y$, $B_z$, $E$) = ($1.08$, $1.2$,
$0.01$, $0.5$, $3.6/\sqrt{4\pi}$, $2/\sqrt{4\pi}$, $0.95$) and the right
state ($1$, $0$, $0$, $0$, $4/\sqrt{4\pi}$, $2/\sqrt{4\pi}$, $1$) with
$B_x=2/\sqrt{4\pi}$ at time $t=0.2$ is plotted in Fig.~2a.
Fast shocks, rotational discontinuities, and  slow shocks
propagate from each side of the contact discontinuity.
The solution of the MHD shock tube test with the left state
($\rho$, $v_x$, $v_y$, $v_z$, $B_y$, $B_z$, $E$) = ($1$, $0$, $0$, $0$,
$6/\sqrt{4\pi}$, $0$, $1$) and the right state ($0.1$, $0$, $2$, $1$,
$1/\sqrt{4\pi}$, $0$, $10$) with $B_x=3/\sqrt{4\pi}$ at time $t=0.035$
is plotted in Fig.~2b.
A fast shock, a rotational discontinuity, and a slow shock propagate from
the left side of the contact discontinuity, while a fast rarefaction,
a rotational discontinuity, and a slow rarefaction propagate to the right.
The rotation across the initial discontinuity of the magnetic field generates
two rotational discontinuities.
As in the previous case, strong fast shocks are resolved within 2-4 cells,
while weak slow shocks take more cells.
With proper stiffening, the rotational discontinuities spread over only
3-5 cells while the contact discontinuity spreads over more cells.

In the third set, handling of magnetosonic structures with vanishing
tangential flow velocity and parallel magnetic field is tested.
A test from DW for magnetosonic shocks is set up with the left state
($\rho$, $v_x$, $v_y$, $v_z$, $B_y$, $B_z$, $E$) = ($0.1$, $50$, $0$, $0$,
$-1/\sqrt{4\pi}$, $-2/\sqrt{4\pi}$, $0.4$) and the right state ($0.1$,
$0$, $0$, $0$, $1/\sqrt{4\pi}$, $2/\sqrt{4\pi}$, $0.2$) and with $B_x=0$.
The solution has a pair of magnetosonic shocks propagating from
a tangential discontinuity and is plotted at time $t=0.01$ in Fig.~3a.
The test for magnetosonic rarefactions are set up with the left state
($\rho$, $v_x$, $v_y$, $v_z$, $B_y$, $B_z$, $E$) = ($1$, $-1$, $0$, $0$,
$1$, $0$, $1$) and the right state ($1$, $1$, $0$, $0$, $1$, $0$, $1$)
and with $B_x=0$.
The solution has only two identical magnetosonic rarefactions
and is plotted at time $t=0.1$ in Fig.~3b.
Magnetosonic shocks are fast shocks with zero parallel field $(B_x=0)$
and they are resolved within 2-4 cells like fast shocks.
The tangential discontinuity, which is a degenerate combination of one contact
discontinuity, two slow structures, and two rotational discontinuities,
spreads typically more than 10 cells.
Even with the stiffening described in \S 2, we could not prevent the
spreading of the tangential discontinuity in numerical calculations.
But the Riemann solutions show that even slight differences from an exact
tangential magnetic field, $B_x = 0$, lead to the formation of the more
complex and expanding set of features already mentioned.

In the fourth set, we test how the MHD-TVD code and the Riemann solver
deal with a special category of fast and slow structures; namely, the so-called
``switch-on''
and ``switch-off'' structures.
The Tangential magnetic field turns on in the region behind switch-on fast
shocks and switch-on slow rarefactions.
On the other hand, it turns off in the region behind switch-off slow
shocks and switch-off fast rarefactions.
The test involving a switch-on fast shock has been set up with the left
state ($\rho$, $v_x$, $v_y$, $v_z$, $B_y$, $B_z$, $E$) = ($1$, $0$, $0$,
$0$, $1$, $0$, $1$) and the right state ($0.2$, $0$, $0$, $0$, $0$, $0$,
$0.1$) and with $B_x=1$.
The switch-on fast shock propagates to the right. Other structures
formed in this test include a fast rarefaction, a slow rarefaction,
a contact discontinuity, and a slow shock.
The test results are plotted at time $t=0.15$ in Fig.~4a, showing that the code
and the Riemann solver handle the switch-on fast shock without any trouble.
The test involving a switch-off fast rarefaction has been set up with
the left state ($\rho$, $v_x$, $v_y$, $v_z$, $B_y$, $B_z$, $E$) =
($0.4$, $-0.66991$, $0.98263$, $0$, $0.0025293$, $0$, $0.52467$) and the
right state ($1$, $0$, $0$, $0$, $1$, $0$, $1$) and with $B_x=1.3$.
It has been designed to generate only a right-moving, switch-off
fast rarefaction with a contact discontinuity with an accuracy (the
ratio of residual strength to pre-rarefaction strength of tangential
magnetic field) better than 0.3\% (see Table 4b).
The test results are plotted at time $t=0.15$ in Fig.~4b.
The numerical calculation shows small, yet noticeable, signatures of the
right-moving slow structure and the left-moving hydrodynamic structure.
The test involving a switch-off slow shock has been set up with the left
state ($\rho$, $v_x$, $v_y$, $v_z$, $B_y$, $B_z$, $E$) = ($0.65$, $0.667$,
$-0.257$, $0$, $0.55$, $0$, $0.5$) and the right state ($1$, $0.4$,
$-0.94$, $0$, $0$, $0$, $0.75$) and with $B_x=0.75$.
The solution of this test has, from left to right, a fast shock, a
switch-off slow shock, a contact discontinuity, and a hydrodynamics shocks.
In the region to the right of the switch-off slow shock hydrodynamic structures
form
with vanishing tangential magnetic field.
The test results are plotted at time time $t=0.15$ in Fig.~4c, showing
good agreement between the numerical calculation and the analytic solution.
Our Riemann solver turns off the tangential magnetic field up to an
accuracy of about 0.04\% as indicated in Table 4c.
The test involving a switch-on slow rarefaction has been set up with
the left state ($\rho$, $v_x$, $v_y$, $v_z$, $B_y$, $B_z$, $E$) =
($1$, $0$, $0$, $0$, $0$, $0$, $1$) and the right state ($0.3$, $0$, $0$,
$1$, $1$, $0$, $0.2$) and with $B_x=0.7$.
The structures formed in this test include a hydrodynamic rarefaction,
a switch-on slow shock, a contact discontinuity, a slow shock, a rotational
discontinuity, and a fast rarefaction.
With the hydrodynamic nature in the region between the two left-moving
rarefactions, no rotational discontinuity is produced on the left side of
the contact discontinuity.
Hence this test has a solution with only one rotational discontinuity,
while two discontinuities are expected in most tests with three-dimensional
field structure (\eg~the tests in Fig.~2).
The test results are plotted at time $t=0.16$ in Fig.~4d.

The final set of tests involves compound structures, whose existence was
first discussed in BW.
The initial setups includes two-dimensional field and velocity structure
in the $x-y$ plane and the sign of the tangential magnetic field $(B_y)$
changes across the initial discontinuity.
The test involving a slow compound structure has been set up with the left
state
($\rho$, $v_x$, $v_y$, $v_z$, $B_y$, $B_z$, $E$) = ($1$, $0$, $0$, $0$,
$1$, $0$, $1$) and the right state ($0.125$, $0$, $0$, $0$, $-1$, $0$, $0.1$)
and with $B_x=0.75$.
It is the same test as in BW, except we set the adiabatic index, $\gamma=5/3$,
instead of $\gamma =2$.
The results are plotted at time $t=0.1$ in Fig.~5a.
Here lines are the solution of the Riemann solver which has a rotational
discontinuity and a slow shock instead of a slow compound structure,
and numerical values of the solution are listed in Table 5a.
Clearly the numerical calculations and the analytic solution produce
different results around the compound structure, even though the
agreement in other structures: two fast rarefactions, a slow shock, and
a contact discontinuity, is acceptable for most purposes.
Similarly, we set up a test involving a fast compound with
the left state ($\rho$, $v_x$, $v_y$, $v_z$, $B_y$, $B_z$, $E$) = ($1$,
$0$, $0$, $0$, $1$, $0$, $1$) and the right state ($0.4$, $0$, $0$, $0$,
$-1$, $0$, $0.4$) and with $B_x=1.3$.
The results are plotted at time $t=0.16$ in Fig.~5b.
Again lines are the solution of the Riemann solver with a rotational
discontinuity and a fast rarefaction instead of a fast compound structure,
and numerical values of the solution are listed in Table 5b.
Further discussion on compound structures will follow in the next
section.

In the above tests with the MHD shock tube problems, the general and detailed
agreement between the numerical calculations and the analytic solutions are
satisfactory, and mostly excellent, except in the tests involving compound
structures.
Especially, the positions of shocks, rarefactions, and discontinuities
agree very well, indicating the both generate the same solutions.
Numerical results show that our code resolves strong shocks (fast, slow,
or magnetosonic) typically within 2-4 cells, while more cells are required
if shocks are weak with small jump in the parallel flow velocity.
Rotational discontinuities are resolved within 3-5 cells with proper
stiffening, and contact discontinuities are resolved within 6-8 cells
without stiffening.
While the stiffening of contact discontinuities make them look sharper in
some cases, it generates spurious numerical oscillations in many cases.
Tangential discontinuities spread over more than 10 cells.
Further work could be done to develop proper stiffening schemes for
contact discontinuities and tangential discontinuities to improve the
ability of the MHD-TVD code to handle these discontinuities.
\section{DISCUSSION}
One test problem that has attracted considerable attention was one
originally presented by BW involving a slow compound structure,
which is related to structures known as ``intermediate'' shocks
(\eg, Wu 1988; Kennel \etal~1990).
This structure reverses the direction of $B_\perp$ and leads to a flow
that passes from super-Alfv\'enic to sub-Alfv\'enic.
Numerical MHD codes, which are always dissipative, seem generally to
find the BW compound structure out of that shock-tube initial condition.
For the compound wave it does not appear necessary that the magnetic field
anywhere exist outside the upstream or downstream field plane.
Thus, it would appear that this structure might be represented as
a ``1 + 1/2 dimensional flow problem''.
However, ideal MHD Riemann solvers that allow rotational discontinuities
will generally find a solution to that shock-tube problem that interprets
the compound structure as a rotational discontinuity followed by
a slow shock, making it a ``1 + 1/2 + 1/2 dimensional'' flow.

It is not our intent to add to the controversy surrounding the reality
of intermediate shocks.
However, in order to test the performance as well as understand the
properties of our code, we have done a high resolution calculation for
the test of the slow compound structure in Fig.~5a with 8192 cells.
The magnified region around the slow compound structure has been plotted
in Fig.~6a.
{}From the plot, the numerical values of several flow quantities on the
both sides of the ``shock'' read: $\rho=0.647$ (density), $p_g=0.484$
(gas pressure), $u=0.963$ (flow velocity in the shock frame), $B_y=0.536$
(tangential magnetic field), $c_f=1.423$ (fast speed), $c_a=0.932$
(Alfv\'en speed), $c_s=0.732$ (slow speed), $a=1.117$ (sound speed) in
the preshock region, $\rho=0.827$, $p_g=0.758$, $u=0.751$, $B_y=-0.0738$,
$c_f=1.241$, $c_a=0.825$, $c_s=0.822$, $a=1.236$ in the postshock region,
and shock velocity in lab frame is 0.284.
In the preshock region the flow velocity is $c_a < u < c_f$ (sub-fast
but super-Alfv\'enic) with Alfv\'enic Mach number $M_a=1.03$, while
in the postshock region $u < c_s$ (sub-slow) with $M_a=0.910$.
Hence, the shock in the numerical calculation with our code should be
considered as a ``2-4'' intermediate shock and the slow compound structure
as a intermediate shock followed by a slow rarefaction.

We have done also a high resolution calculation for the test of the fast
compound structure in Fig.~5b with 8192 cells and the magnified region
around the fast compound structure has been plotted in Fig.~6b.
The numerical values in the plot of several flow quantities on the
both sides of the ``shock'' read: $\rho=0.6820$, $p_g=0.5290$, $u=1.617$,\
$B_y=0.0555$, $c_f=1.577$, $c_a=1.574$, $c_s=1.135$, $a=1.137$ in the
preshock region, $\rho=0.7361$, $p_g=0.6031$, $u=1.498$, $B_y=-0.3442$,
$c_f=1.622$, $c_a=1.515$, $c_s=1.092$, $a=1.169$ in the postshock region,
and shock velocity in lab frame is 0.935.
In the preshock region the flow velocity is $u > c_f$ (super-fast) with
$M_a=1.027$, while in the postshock region $c_s < u < c_a $ (sub-Alfv\'enic
but super-slow) with $M_a=0.989$.
Hence, the shock in the numerical calculation with our code should be
considered as a ``1-3'' intermediate shock and the fast compound structure
as a intermediate shock preceded by a fast rarefaction.

%
\heading{Acknowledgments}
The work by DR was supported in part by David and Lucille Packard
Foundation Fellowship through Jeremy Goodman at Princeton University
and in part by Non-Directed Research Fund of Korea Research
Foundation 1993 at Chungnam National University. Work by TWJ was supported in
part
by NASA
through grant NAGW-2548, the NSF through grant AST-9100486, by a travel grant
from the University of Minnesota International Programs Office and
by the University of Minnesota Supercomputer Institute (UMSI). The
UMSI also generously contributed support for travel by DR.
\vfill\eject
\heading{REFERENCES}
\journal{Brio, M., \& Wu, C.~C.}{1988}{J.~ Comp.~ Phys.}{75}{500}
\journal{Colella, P., \& Woodward, P.~R.}{1984}{J.~ Comp.~ Phys.}{54}{174}
\infuture{Dai W., \& Woodward P.~R.}{1994}{J.~of Comput.~Phys.}{in press}
\journal{DeVore, C.~R.}{1991}{J.~of Comput.~Phys.}{92}{142}
\journal{Harten, A.}{1983}{J.~ Comp.~ Phys.}{49}{357}
\book{Jeffrey, A.}{1966}{Magnetohydrodynamics}{London}{Oliver and Boyd}
\book{Jeffrey A., \& Taniuti, T.}{1964}{Nonlinear Waves Propagation}
{New York}{Academic Press}
\journal{Kennel, C.~F., Blandford, R. D. \& Wu, C. C.}{1990}{Phys. Fluids
B}{2}{253}
\privcom{Lind, K. R., Payne, D. G. \& Meier, D. L.}{1991}{preprint}
\book{Parker, E.~N.}{1979}{Cosmical Magnetic Fields: Their Origin and Their
Activity}{Oxford}{Oxford University Press}
\journal{Roe, P.~L.}{1981}{J.~ Comp.~ Phys.}{43}{357}
\book{Shu, F.~H.}{1992}{The Physics of Astrophysics, Volume II: Gad Dynamics}
{Mill Valley}{University Science Books}
\journal{Stone, J. M., Hawley, J. F., Evans, C. R. \& Norman, M.
L.}{1992}{ApJ}{388}{415}
\journal{Stone, J.~M., \& Norman, M.~L.}{1992}{ApJS}{80}{791}
\journal{Wu, C.~C.}{1988}{J.~Geophys.~Research}{93}{987}
\journal{Zachary A.~L., \& Colella P.}{1992}{J.~of Comput.~Phys.}{99}{341}
\vfill\eject
\heading{FIGURE CAPTIONS}
\figcap{1a}{Solution of the MHD shock tube test with the left state
($\rho$, $v_x$, $v_y$, $v_z$, $B_y$, $B_z$, $E$) = ($1$, $10$, $0$, $0$,
$5/\sqrt{4\pi}$, $0$, $20$) and the right state ($1$, $-10$, $0$, $0$,
$5/\sqrt{4\pi}$, $0$, $1$) with $B_x=5/\sqrt{4\pi}$ and $\gamma=5/3$ at
time $t=0.08$ (test in DW table 7).
Dots are the result of a numerical calculation with the MHD-TVD code
described in \S 2 using 512 cells and a Courant constant of 0.8.
Lines are the result with the nonlinear Riemann solver described in \S 3.
Plots show from left to right (1) fast shock, (2) slow rarefaction,
(3) contact discontinuity, (4) slow shock, and (5) fast shock.}

\figcap{1b}{Solution of the MHD shock tube test with the left state
($\rho$, $v_x$, $v_y$, $v_z$, $B_y$, $B_z$, $E$) = ($1$, $0$, $0$, $0$,
$5/\sqrt{4\pi}$, $0$, $1$) and the right state ($0.1$, $0$, $0$, $0$,
$2/\sqrt{4\pi}$, $0$, $10$) with $B_x=3/\sqrt{4\pi}$ and $\gamma=5/3$ at
time $t=0.03$ (test in DW table 3a).
Dots are the result of a numerical calculation with the MHD-TVD code
described in \S 2 using 512 cells and a Courant constant of 0.8.
Lines are the result with the nonlinear Riemann solver described in \S 3.
Plots show from left to right (1) fast shock, (2) slow shock,
(3) contact discontinuity, (4) slow rarefaction, and (5) fast rarefaction.}

\figcap{2a}{Solution of the MHD shock tube test with the left state
($\rho$, $v_x$, $v_y$, $v_z$, $B_y$, $B_z$, $E$) = ($1.08$, $1.2$,
$0.01$, $0.5$, $3.6/\sqrt{4\pi}$, $2/\sqrt{4\pi}$, $0.95$) and the right
state ($1$, $0$, $0$, $0$, $4/\sqrt{4\pi}$, $2/\sqrt{4\pi}$, $1$) with
$B_x=2/\sqrt{4\pi}$ and $\gamma=5/3$ at time $t=0.2$ (test in DW table 1a).
Dots are the result of a numerical calculation with the MHD-TVD code
described in \S 2 using 512 cells and a Courant constant of 0.8.
Lines are the result with the nonlinear Riemann solver described in \S 3.
Plots show from left to right (1) fast shock, (2) rotational discontinuity,
(3) slow shock, (4) contact discontinuity, (5) slow shock, (6) rotational
discontinuity, and (7) fast shock.}

\figcap{2b}{Solution of the MHD shock tube test with the left state
($\rho$, $v_x$, $v_y$, $v_z$, $B_y$, $B_z$, $E$) = ($1$, $0$, $0$, $0$,
$6/\sqrt{4\pi}$, $0$, $1$) and the right state ($0.1$, $0$, $2$, $1$,
$1/\sqrt{4\pi}$, $0$, $10$) with $B_x=3/\sqrt{4\pi}$ and $\gamma=5/3$ at
time $t=0.035$ (test in DW table 5a).
Dots are the result of a numerical calculation with the MHD-TVD code
described in \S 2 using 512 cells and a Courant constant of 0.8.
Lines are the result with the nonlinear Riemann solver described in \S 3.
Plots show from left to right (1) fast shock, (2) rotational discontinuity,
(3) slow shock, (4) contact discontinuity, (5) slow rarefaction, (6) rotational
discontinuity, and (7) fast rarefaction.}

\figcap{3a}{Solution of the MHD shock tube test with the left state
($\rho$, $v_x$, $v_y$, $v_z$, $B_y$, $B_z$, $E$) = ($0.1$, $50$, $0$, $0$,
$-1/\sqrt{4\pi}$, $-2/\sqrt{4\pi}$, $0.4$) and the right state ($0.1$,
$0$, $0$, $0$, $1/\sqrt{4\pi}$, $2/\sqrt{4\pi}$, $0.2$) with $B_x=0$ and
$\gamma=5/3$ at time $t=0.01$ (test in DW table 2a).
Dots are the result of a numerical calculation with the MHD-TVD code
described in \S 2 using 512 cells and a Courant constant of 0.8.
Lines are the result with the nonlinear Riemann solver described in \S 3.
Plots show from left to right (1) magnetosonic shock, (2) tangential
discontinuity, and (3) magnetosonic shock.}

\figcap{3b}{Solution of the MHD shock tube test with the left state
($\rho$, $v_x$, $v_y$, $v_z$, $B_y$, $B_z$, $E$) = ($1$, $-1$, $0$, $0$,
$1$, $0$, $1$) and the right state ($1$, $1$, $0$, $0$, $1$, $0$, $1$)
with $B_x=0$ and $\gamma=5/3$ at time $t=0.1$.
Dots are the result of a numerical calculation with the MHD-TVD code
described in \S 2 using 512 cells and a Courant constant of 0.8.
Lines are the result with the nonlinear Riemann solver described in \S 3.
Plots show from left to right (1) magnetosonic rarefaction, and (2)
magnetosonic rarefaction.}

\figcap{4a}{Solution of the MHD shock tube test with the left state
($\rho$, $v_x$, $v_y$, $v_z$, $B_y$, $B_z$, $E$) = ($1$, $0$, $0$, $0$,
$1$, $0$, $1$) and the right state ($0.2$, $0$, $0$, $0$, $0$, $0$, $0.1$)
with $B_x=1$ and $\gamma=5/3$ at time $t=0.15$.
Dots are the result of a numerical calculation with the MHD-TVD code
described in \S 2 using 512 cells and a Courant constant of 0.8.
Lines are the result with the nonlinear Riemann solver described in \S 3.
Plots show from left to right (1) fast rarefaction, (2) slow rarefaction,
(3) contact discontinuity, (4) slow shock, and (5) switch-on fast shock.}

\figcap{4b}{Solution of the MHD shock tube test with the left state
($\rho$, $v_x$, $v_y$, $v_z$, $B_y$, $B_z$, $E$) = ($0.4$, $-0.66991$,
$0.98263$, $0$, $0.0025293$, $0$, $0.52467$) and the right state ($1$,
$0$, $0$, $0$, $1$, $0$, $1$) with $B_x=1.3$ and $\gamma=5/3$ at
time $t=0.15$.
Dots are the result of a numerical calculation with the MHD-TVD code
described in \S 2 using 512 cells and a Courant constant of 0.8.
Lines are the result with the nonlinear Riemann solver described in \S 3.
Plots show from left to right (1) contact discontinuity, and (2)
switch-off fast rarefaction.}

\figcap{4c}{Solution of the MHD shock tube test with the left state
($\rho$, $v_x$, $v_y$, $v_z$, $B_y$, $B_z$, $E$) = ($0.65$, $0.667$,
$-0.257$, $0$, $0.55$, $0$, $0.5$) and the right state ($1$, $0.4$, $-0.94$,
$0$, $0$, $0$, $0.75$) with $B_x=0.75$ and $\gamma=5/3$ at time $t=0.15$.
Dots are the result of a numerical calculation with the MHD-TVD code
described in \S 2 using 512 cells and a Courant constant of 0.8.
Lines are the result with the nonlinear Riemann solver described in \S 3.
Plots show from left to right (1) fast shock, (2) switch-off slow shock,
(3) contact discontinuity, and (4) hydrodynamic shock.}

\figcap{4d}{Solution of the MHD shock tube test with the left state
($\rho$, $v_x$, $v_y$, $v_z$, $B_y$, $B_z$, $E$) = ($1$, $0$, $0$, $0$,
$0$, $0$, $1$) and the right state ($0.3$, $0$, $0$, $1$, $1$, $0$, $0.2$)
with $B_x=0.7$ and $\gamma=5/3$ at time $t=0.16$.
Dots are the result of a numerical calculation with the MHD-TVD code
described in \S 2 using 512 cells and a Courant constant of 0.8.
Lines are the result with the nonlinear Riemann solver described in \S 3.
Plots show from left to right (1) hydrodynamic rarefaction, (2) switch-on
slow rarefaction, (3) contact discontinuity, (4) slow shock, (5) rotational
discontinuity, and (6) fast rarefaction.}

\figcap{5a}{Solution of the MHD shock tube test with the left state
($\rho$, $v_x$, $v_y$, $v_z$, $B_y$, $B_z$, $E$) = ($1$, $0$, $0$, $0$,
$1$, $0$, $1$) and the right state ($0.125$, $0$, $0$, $0$, $-1$, $0$, $0.1$)
with $B_x=0.75$ and $\gamma=5/3$ at time $t=0.1$ (test in BW).
Dots are the result of a numerical calculation with the MHD-TVD code
described in \S 2 using 512 cells and a Courant constant of 0.8.
Lines are the result with the nonlinear Riemann solver described in \S 3.
Plots show from left to right (1) fast rarefaction, (2) slow compound,
(3) contact discontinuity, (4) slow shock, and (5) fast rarefaction.}

\figcap{5b}{Solution of the MHD shock tube test with the left state
($\rho$, $v_x$, $v_y$, $v_z$, $B_y$, $B_z$, $E$) = ($1$, $0$, $0$, $0$,
$1$, $0$, $1$) and the right state ($0.4$, $0$, $0$, $0$, $-1$, $0$, $0.4$)
with $B_x=1.3$ and $\gamma=5/3$ at time $t=0.16$.
Dots are the result of a numerical calculation with the MHD-TVD code
described in \S 2 using 512 cells and a Courant constant of 0.8.
Lines are the result with the nonlinear Riemann solver described in \S 3.
Plots show from left to right (1) fast compound, (2) slow shock, (3)
contact discontinuity, (4) slow shock, and (5) fast rarefaction.}

\figcap{6a}{Same slow compound structure in the MHD shock tube test as
that in Fig.~5a.
The calculation has been done with the MHD-TVD code described in \S 2
using 8192 cells and only the region around the slow compound structure
has been plotted.
Plots show that the slow compound structure from the numerical calculation
is composed of a ``2-4'' intermediate shock followed by a slow rarefaction.}

\figcap{6b}{Same fast compound structure in the MHD shock tube test as
that in Fig.~5b.
The calculation has been done with the MHD-TVD code described in \S 2
using 8192 cells and only the region around the fast compound structure
has been plotted.
Plots show that the fast compound structure from the numerical calculation
is composed of a ``1-3'' intermediate shock preceded by a fast rarefaction.}
\bye